\documentclass[10pt,journal,compsoc]{IEEEtran}
\usepackage{float}
\usepackage{url}
\usepackage{graphicx}
\usepackage{longtable}
\usepackage{multirow}
\usepackage{colortbl}
\usepackage[table,xcdraw]{xcolor}
\usepackage{array}
\usepackage{array,booktabs,enumitem}
\newcolumntype{P}[1]{>{\endgraf\vspace*{-\baselineskip}}p{#1}}
\ifCLASSOPTIONcompsoc
  \usepackage[nocompress]{cite}
\else
  \usepackage{cite}
\fi
\ifCLASSINFOpdf
\else
\fi

\usepackage{pdflscape}
\usepackage{afterpage}
\usepackage{float}
\usepackage{xurl}

\usepackage{etoolbox}
\usepackage{xspace}
\usepackage{xcolor}
\usepackage[all]{xy}
\usepackage{hyperref}
\usepackage{cleveref}
\usepackage{tcolorbox}
\tcbuselibrary{breakable}
\usepackage{lipsum,multicol}
\usepackage{multirow}
\usepackage{bigstrut}
\usepackage{bigdelim}
\usepackage{longtable}
\usepackage{scrextend}
\usepackage[normalem]{ulem}
\usepackage{ragged2e}

\begin{document}
\newtoggle{draft}
\toggletrue{draft}
\iftoggle{draft}
{
	\newcommand{\jon}[1]{{\leavevmode\color{red}{Jon: #1}\xspace}}
	\newcommand{\waqar}[1]{{\leavevmode\color{blue}{Waqar: #1}\xspace}}
	\newcommand{\rifat}[1]{{\leavevmode\color{magenta}{Rifat: #1}\xspace}}
	\newcommand{\harsha}[1]{{\leavevmode\color{orange}{Harsha: #1}\xspace}}
	\newcommand{\arif}[1]{{\leavevmode\color{magenta}{Arif: #1}\xspace}}
	\newcommand{\mojtaba}[1]{{\leavevmode\color{brown}{Mojtaba: #1}\xspace}}
	\newcommand{\rashina}[1]{{\leavevmode\color{violet}{Rashina: #1}\xspace}}
	\newcommand{\gillian}[1]{{\leavevmode\color{cyan}{Gillian: #1}\xspace}}

}
{
	\newcommand{\jon}[1]{}
	\newcommand{\waqar}[1]{}
	\newcommand{\rifat}[1]{}
	\newcommand{\harsha}[1]{}
	\newcommand{\arif}[1]{}
	\newcommand{\mojtaba}[1]{}
	\newcommand{\rashina}[1]{}
	\newcommand{\gillian}[1]{}
}

%

\title{The Impact of Considering Human Values during Requirements Engineering Activities}

\author{Harsha~Perera,~\IEEEmembership{}
        Rashina~Hoda,~\IEEEmembership{}
        Rifat~Ara~Shams,~\IEEEmembership{}\\
        Arif~Nurwidyantoro,~\IEEEmembership{}
        Mojtaba~Shahin,~\IEEEmembership{}
        Waqar~Hussain,~\IEEEmembership{}
        and~Jon~Whittle~\IEEEmembership{}
\IEEEcompsocitemizethanks{\IEEEcompsocthanksitem 
Harsha Perera, Rashina Hoda, Rifat Ara Shams, Arif Nurwidyantoro, Mojtaba Shahin, and Waqar Hussain are with the Faculty of Information Technology, Monash University, Australia.
\protect\\
E-mail: \{harsha.perera,~rashina.hoda,~rifat.shams,~arif.nurwidyantoro, ~mojtaba.shahin,~waqar.hussain\}@monash.edu}\\
\IEEEcompsocitemizethanks{\IEEEcompsocthanksitem 
Jon Whittle is with CSIRO's Data61, Australia.
\protect\\
E-mail: Jon.Whittle@data61.csiro.au
}

\thanks{Manuscript submitted to IEEE Transactions on Software Engineering (2021)}
}

%
%

\markboth{IEEE Transactions on Software Engineering}%
{Shell \MakeLowercase{\textit{Perera et al.}}: Bare Demo of IEEEtran.cls for Computer Society Journals}
%



\IEEEtitleabstractindextext{%
\begin{abstract}
\justifying
Human values, or \textit{what people hold important in their life}, such as freedom, fairness, and social responsibility, often remain unnoticed and unattended during software development. Ignoring values can lead to values violations in software that can result in financial losses, reputation damage, and widespread social and legal implications. However, embedding human values in software is not only non-trivial but also generally an unclear process. Commencing as early as during the Requirements Engineering (RE) activities promises to ensure fit-for-purpose and quality software products that adhere to human values. But what is the impact of considering human values explicitly during early RE activities? To answer this question, we conducted a scenario-based survey where 56 software practitioners contextualised requirements analysis towards a proposed mobile application for the homeless and suggested values-laden software features accordingly. The suggested features were qualitatively analysed. Results show that explicit considerations of values can 
help practitioners identify applicable values, associate  purpose  with  the  features  they  develop, think \textit{outside-the-box}, and build connections  between  software  features  and  human values. 
Finally, drawing from the results and experiences of this study, we propose a scenario-based values elicitation process -- a simple four-step takeaway as a practical implication of this study.   
\end{abstract}

\begin{IEEEkeywords}
Human Values, Requirements Engineering, Survey, Scenario
\end{IEEEkeywords}}

\maketitle

\IEEEdisplaynontitleabstractindextext

%
\IEEEpeerreviewmaketitle

\section{Introduction}
\label{sec:Introduction}

Software is an inextricable part of our lives and social fabric. People expect software to demonstrate and respect human values -- \textit{what people hold important in their life} \cite{schwartz2012overview} -- such as social justice, freedom, independence, fairness, accessibility, and tradition. Unsurprisingly, values violations through software applications often end up creating undesired consequences such as financial losses~\cite{neate_2018}, reputation damages~\cite{Galhotra2017}, and even loss of lives~\cite{molly.2019}. For example, Facebook (now renamed as Meta) recently changed WhatsApp's term and conditions, leaving no choice for users but to grant access of their personal data including phone number and behaviour to Facebook or lose their WhatsApp account~\cite{best2021}. People accused Facebook of violating their \textit{trust} and \textit{freedom to choose}, and this change led millions of WhatsApp users to migrate to alternative messaging apps, such as Telegram and Signal~\cite{best2021}. 
In a more severe example, ``Blue Whale Challenge'' -- a game conducted through social media apps was responsible for the death
of 153 teenagers around the world~\cite{gowda2019blue}. The game presented its players with 50 tasks in 50 days, and the 50\textsuperscript{th} task was to take your own life~\cite{mukhra2019blue}, possibly violating all human values imaginable.


Engineering human values into software is challenging due to their ill-defined nature in the software context~\cite{Mougouei2018}. Ferrario et al. argue that `values embedded into software are often invisible and taken for granted'~\cite{Ferrario2014}. Thew and Sutcliffe argue that stakeholder values, motivations and emotions are not explicitly addressed in the requirements processes~\cite{thew2018value}. Further, Harbers and others propose ``\textit{explicitly identifying and considering stakeholder values during requirements elicitation, identification and analysis will lead to software that better supports human values}''~\cite{harbers2015embedding}. There are commendable recent efforts in Requirements Engineering (RE) that support the effort of explicit consideration of human values in software such as values based requirements engineering (VBRE)~\cite{thew2018value},  HuValue -- a values based design tool~\cite{kheirandish2019huvalue}, and the value story workshop~\cite{harbers2015embedding}. All of these studies seem to hold a common assumption that \textit{explicitly considering human values in RE would make software better aligned with values}. In this study, we challenge this assumption by examining it with real-world software practitioners who engage in RE activities, with the following research question:


\begin{quote}
    \textit{\textbf{What is the impact of considering human values explicitly in the early requirements engineering activities?}} 
\end{quote}
To address this research question, we designed a scenario-based survey, where a scenario was presented to the respondents and while explicitly thinking about human values, they were asked to suggest features that would satisfy the requirements given in the scenario. We discuss the survey design and flow in detail in Section~\ref{Method_Survey_design}, as well as our reflections on its design and outcomes in Section \ref{sec:reflections}.

Given the detailed nature of the scenario-based survey, taking an average of ~30 minutes to complete, we were pleased to receive responses from 56 software practitioners who engaged in RE activities on a regular basis. Apart from demographics and values familiarity details, the responses mainly consisted of the descriptions of the suggested features with due consideration of human values. We used these feature descriptions as the primary unit of qualitative analysis. Using socio-technical grounded theory (STGT) \textit{for data analysis} \cite{hoda2021STGT}, the characteristics of the suggested feature were rigorously analysed to inform us of the impact of considering human values explicitly during RE activities in terms of helping practitioners deign more values-focused software features. 

This research makes the following key contributions:


\begin{itemize}
    \item Providing empirical evidence of the impact of considering human values during RE activities in improving values identification and mapping. 
    \item Promoting scenario-based thinking as an effective tool for operationlizing human values in RE.
    \item Introducing a four-step \textit{scenario-based values elicitation} process as a practical takeaway for RE practitioners to consider human values in their day-to-day requirements analysis work and for researchers to adapt and use in similar contexts.
    \item Introducing scenario-based surveys as an effective and flexible research tool for addressing complex research questions that require experiential evidence and flexibility under physical and time constraints such as those imposed by the Covid-19 pandemic related  work-from-home conditions at scale.
\end{itemize}


\begin{figure*}[t]
    \centering
    \includegraphics[width=1\textwidth]{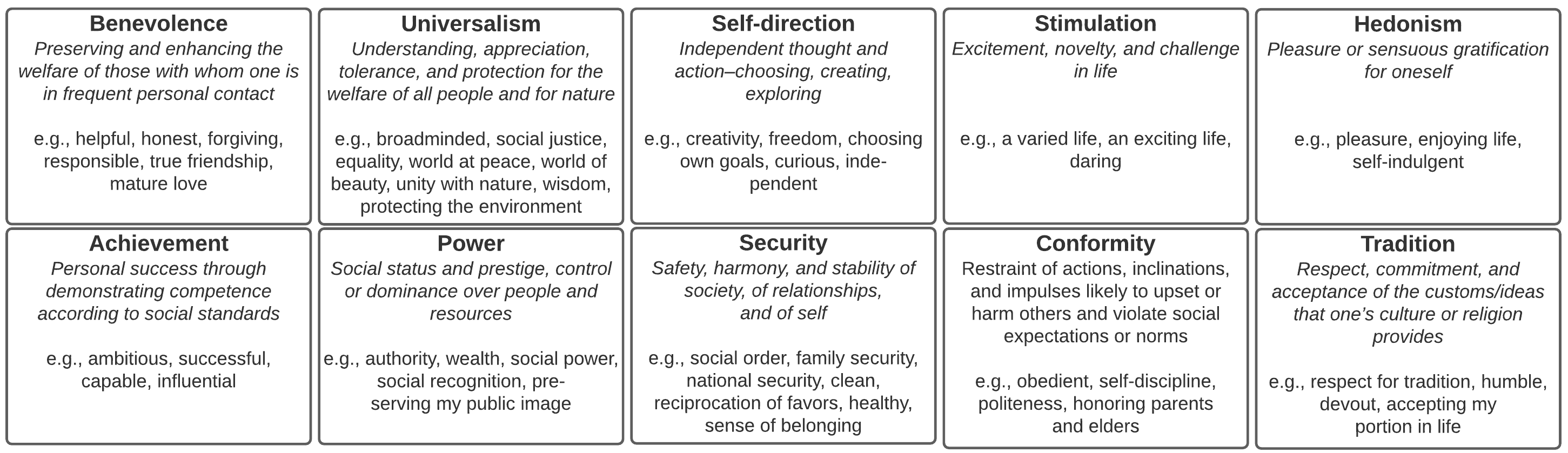}
	\caption{Values categories, definitions and examples of included individual values as in Schwartz's theory of basic human values ~\cite{schwartz1992universals}}
	\label{fig:values}
\end{figure*}
\section{Related Work}
\label{sec:RelatedWork}

\begin{figure*}[t]
    \centering
    \includegraphics[width=1\textwidth]{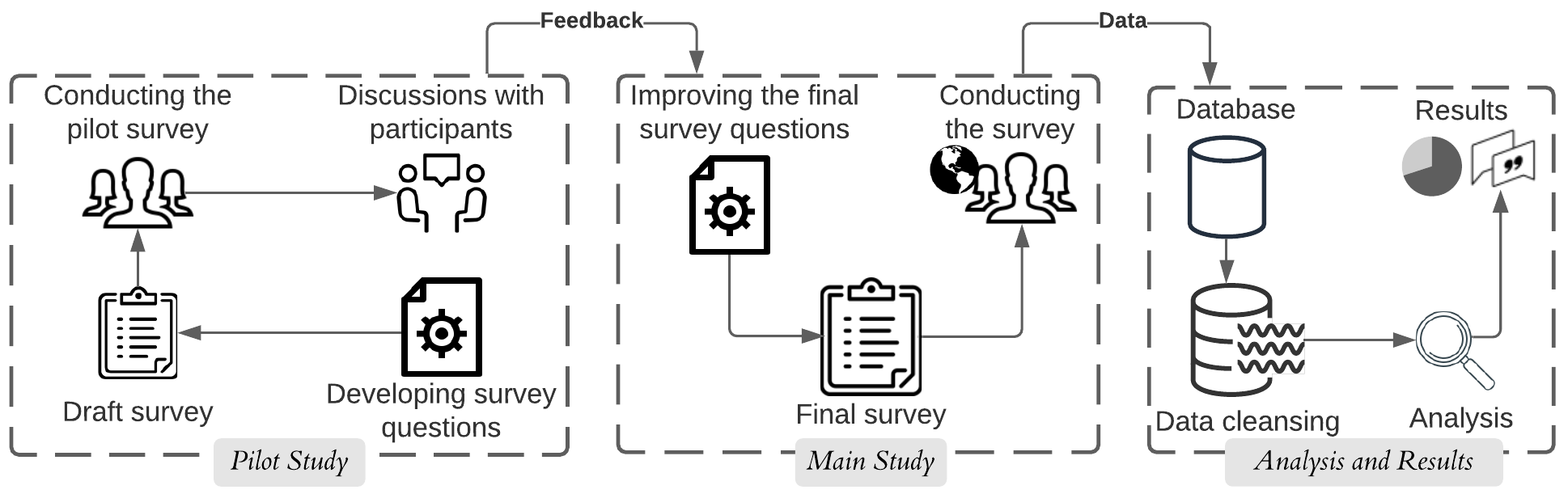}
	\caption{Research methodology overview including pilot study, main study (scenario-based survey), analysis and presentation of results.}
	\label{fig:methodology}
\end{figure*}

\subsection{Human Values}
\label{subsec:HumanValues}

\textit{\textbf{Human Values Definition:}} Human values are defined by Schwartz as \textit{standards that we use to judge the appropriateness of attitudes, traits or virtues} \cite{schwartz2012overview}. Meanwhile, seven different definitions of human values are summarized as \textit{``guiding principles of what people consider important in life''} \cite{cheng2010developing}. 
In software engineering contexts, human values represent the characteristics of software that are considered as important for the stakeholders \cite{NURWIDYANTORO2022106731}. They contain but are not limited to values of ethical importance, often known as ethics. \newline 
\textit{\textbf{Human Values Representation:}}
Since the 1950s, social scientists have been searching for the most useful way to conceptualize basic human values~\cite{schwartz2007basic}. 
In 1973, Rokeach captured 36 human values and organized them into two categories as terminal values and instrumental values~\cite{rokeach1973nature}. In 1980, Hofstede divided values into two categories, desired (what people actually desire) and desirable (what people think ought to be desired) \cite{hofstede1980culture}. In 1992, Schwartz introduced the theory of basic human values, which is assessed across 82 countries~\cite{schwartz2012overview}. It identified ten motivationally-distinct values categories and measured them using 58 distinct values~\cite{schwartz1992universals, schwartz2012overview}. 
Although there are many more 
classifications for human values \cite{cheng2010developing}, in this research, we use Schwartz's theory, which is the most cited and widely applied classification not only in the social sciences but also in other disciplines~\cite{thew2018value, Ferrario2014}. 
Figure~\ref{fig:values} depicts Schwartz's values categories, their definitions and individual values.


\subsection{Human Values in Technology Design}
\label{subsec:HumanValueTechnology}

Since the 1970s, research on human values in technology design and development has been going on \cite{van2015design}. Attempts to consider human values during technology design and development have been of interest particularly in the field of Human-Computer Interaction (HCI). The first attempt came from Batya Friedman by proposing an approach called Value-Sensitive Design (VSD) to elicit values and integrate them in technology design \cite{van2015design}. According to Friedman et al., ``Value Sensitive Design is a theoretically grounded approach to the design of technology that accounts for human values in a principled and comprehensive manner throughout the design process"~\cite{friedman2008value}. Friedman et al. also explored the conceptual, empirical, and technical aspects of VSD and provided suggestions accordingly to use VSD~\cite{friedman2008value}. However, VSD is often being questioned for limiting to values with ethical or moral importance~\cite{davis2015value,ferrario2016values}. While morality or ethics may judge right from wrong, values do not necessarily have an ethical import all the time (authority, ambitious, capable, pleasure for example). 
Considering only a subset of human values makes VSD incomplete to address the challenge of integrating human values into software~\cite{winter2019advancing}. Moreover, translating identified human values into corresponding design features in the system is an underdeveloped activity in VSD \cite{harbers2015embedding,detweiler2011principles}.



\begin{figure*}[htb]
    \centering
    \includegraphics[width=1\textwidth]{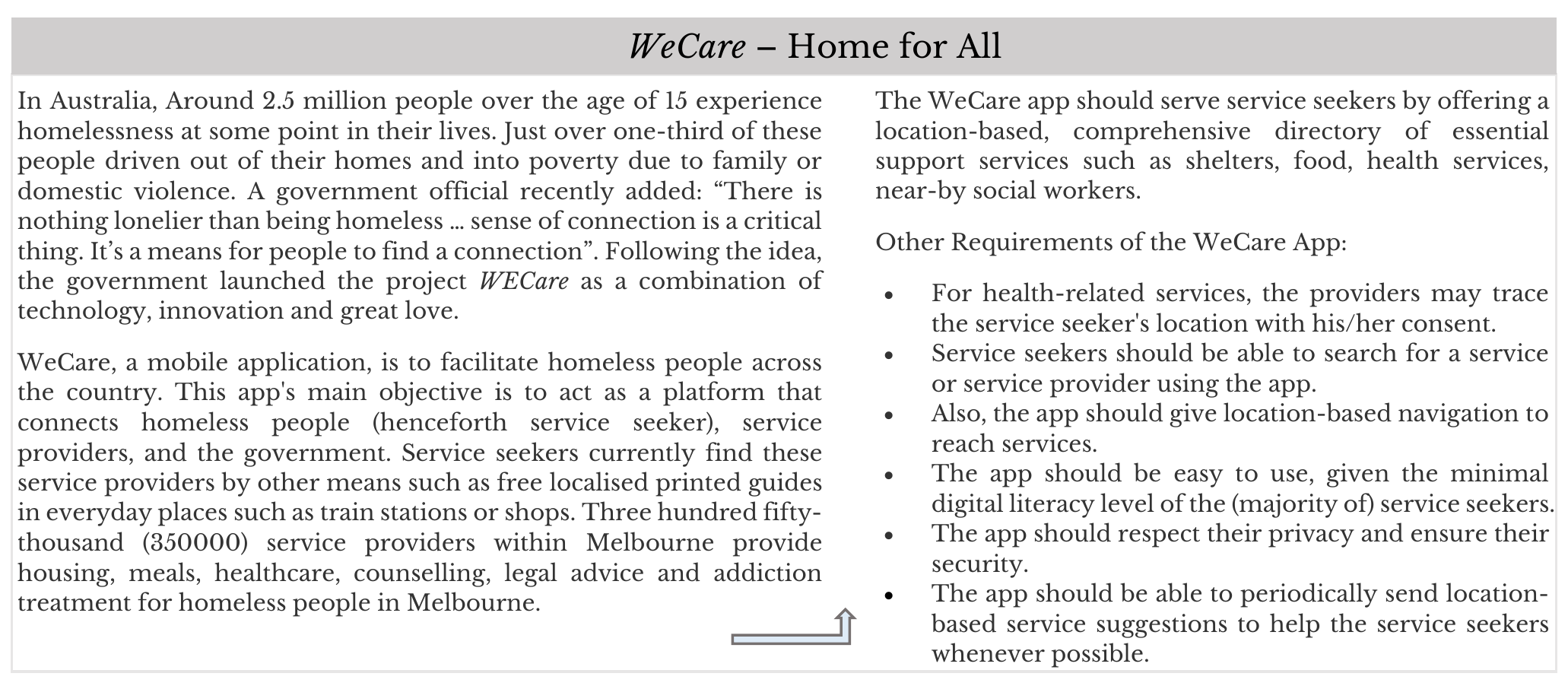}
	\caption{WeCare App scenario presented in the survey}
	\label{fig:wecare}
\end{figure*}

\subsection{Human Values in Software Engineering}
\label{subsec:HumanValueSE}

Recent studies proposed several approaches to support the integration of human values suitable for software engineering. Some studies focused on increasing the awareness of values during software development. For example, Mougouei et al. explained the importance of addressing human values in software, identified the research gap of measuring human values in SE and proposed a research roadmap to address human values in SE \cite{mougouei2018operationalizing}. Ferrario et al. proposed an approach called Values-First SE that uses action research techniques to map requirements to values and reflect the values from users feedback \cite{ferrario2016values}. 
Another study identified intervention points (e.g., artefacts, ceremonies, practices) to integrate values in a well-known agile framework \cite{hussain2021can}. Winter et al. designed and developed a value measurement tool named Values Q-Sort to investigate values at system, personal, and instantiation levels of SE and to establish value relations \cite{winter2018measuring}.

There are also a few studies to address values in different phases of software development life-cycle (SDLC). 
For example, a recent study proposed a dashboard tool for software repositories to address values-related issues during SDLC \cite{nurwidyantoro2021towards}. Meanwhile, other studies attempt to address values in a specific phase of the SDLC, such as requirements and design, as presented in the following subsections. 

\subsubsection{Human Values in Software Design}
\label{subsubsec:HumanValueSDesign}

Design phase is considered as a potential place to \textit{``realize values''} \cite{Steen2012}. For this reason, several studies proposed approaches to support values consideration in the design process of technology. One of the approaches was used by Aldewereld et al. to propose a framework that creates explicit links between the values and the corresponding architectural and design decisions to maintain the values during development \cite{aldewereld2015design}. This framework is called Value-Sensitive Software Development (VSSD) that used `Design for Values' approach \cite{aldewereld2015design}. In another study, Hussain et al. proposed a framework to consider human values in design patterns \cite{hussain2018integrating}.


\subsubsection{Human Values in Requirements}
\label{subsubsec:HumanValueRE}

Failure to distinguish between user and system requirements may lead to soft issues in RE, such as politics and people’s feelings, motivations and values~\cite{thew2018value}. However, RE offers relatively little guidance to deal with them, and human values are rarely considered among soft issues compared to quality aspects of software such as privacy or security~\cite{harbers2015embedding}. Detweiler and Harbers explain this ignorance as `\textit{thinking about values is not common practice in RE}'~\cite{detweiler2014value}. 

However, to address human values in software, it is necessary to capture them in the requirements during the RE activities. Here, we acknowledge, recent, but isolated RE approaches that recognised human values explicitly in their research.  Value-Based Requirements Engineering (VBRE)~\cite{thew2018value}, uses stakeholders' values, motivations, and emotions (VME) to elicit and analyze soft issues of the software. However, VBRE identifies the RE process management implications that values bring about, rather than providing proper guidance to convert identified values to features of the system~\cite{harbers2015embedding}. Duboc et al. considered non-technical aspects of software, such as ethics, power, politics, and values, by utilizing critical system thinking in the early requirements engineering process \cite{duboc2019critical}. Another couple of studies suggest two different model languages to model emotions~\cite{migueis2019towards, pedell2015don}. A more recent effort in RE to address human values is HuValue tool that supports designers in considering human values in their design~\cite{kheirandish2019huvalue}. 

While each of these research has its limitations such as VBRS provides less guidance to convert identified values to features of the system~\cite{harbers2015embedding}, they hold a common assumption that explicitly considering  human  values in RE would make software better aligned with values. While each of these research has its limitations, they hold a common assumption that explicitly considering human values in RE would make software better aligned with values, however, it requires RE research to prove such impact. To the best of our knowledge, there is no research to examine this assumption effectively with empirical evidence. Therefore, this study aims to investigate the impact of considering human values explicitly in RE activities on software features.

\section{Methodology}
\label{sec:Methodology}

\begin{figure*}[t]
    \centering
    \includegraphics[width=1\textwidth]{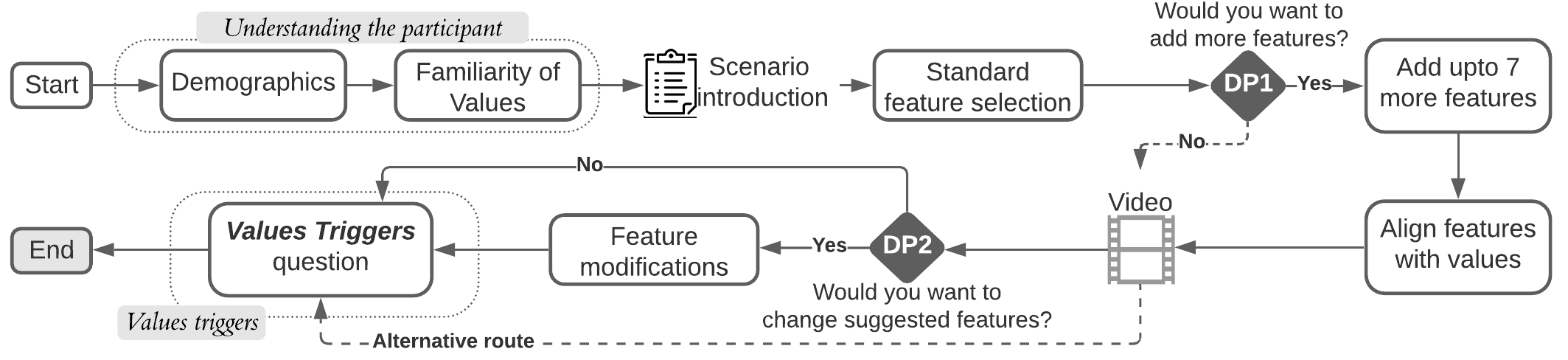}
	\caption{Survey design and flow paths (DP: Decision Point)}
	\label{fig:surveyflow}
\end{figure*}
\subsection{Survey Design} 
\label{Method_Survey_design}

We conducted scenario-based survey research to study the impact of explicit consideration of human values in the early Requirements Engineering (RE) activities (e.g., requirements analysis) on software design. The scenario-based survey was in-depth and involved a hypothetical case to consider. The approach was used to overcome the challenges of conducting in-person workshops during the \textit{Covid-19} pandemic situation. Using a survey, we wanted to reach a broader population of software practitioners involved in RE activities. The overall methodology consists of three stages, namely, a pilot study, a main study, and data analysis (see Fig.\ref{fig:methodology}).

First, as a preparation, we set up the survey goals during, designed the research flow, and drafted the scenario along with the initial survey questionnaire. We started the pilot study after obtaining approval from the Monash University Human Research Ethics Committee (MUHREC) (project number 25278).

\textit{\textbf{Pilot Study:}} 
We conducted a pilot study with four industry practitioners selected from our contacts to assess our survey design and in particular, the clarity of the survey questions and the comprehensibility of the scenario. The pilot participants were asked to use the `\textit{think-aloud}' technique in which they voice-recorded their feedback to the survey questions (if any) while doing the survey \cite{nielsen2002getting}. Then we carried out short discussions with pilot participants to elicit further suggestions. Based on the analysis of the think-aloud voice recordings and the researcher's notes of the discussions, we improved the phrasing of a few questions and added external links to access the Schwartz's model in the question descriptions. Further, we updated several exit points of the survey and streamlined the survey logic. However, none of these changes affected the principle design or the intention of the survey. Rather, they served to improve the flow, comprehensibility, and information needs of the respondents. Once finalised, we continued to conduct the main study. The following subsections discuss the final survey design, data collection, and data analysis approaches in detail.

\subsubsection{Understanding the Participants}
The first section of the survey collected demographic information about the participants, including their job roles and experience in the software industry (see Fig.~\ref{fig:surveyflow}). Further, the section questioned to what extent they elicit, analyse, prioritise and design software requirements as a part of their job. Finally, the section evaluated their level of familiarity with human values. Schwartz's theory of basic human values was used to define human values, and in the online survey platform, the cards shown in Fig.\ref{fig:values} were displayed as clickable areas to pick the values that they were already familiar with. This served the dual purposes of gauging the participants' prior knowledge of human values and introducing them to (or reminding them of) the Schwartz's model.


\subsubsection{Scenario Design: WeCare App}


The survey was developed around a hypothetical scenario of a proposed mobile app (\textit{WeCare}) for homeless people in Australia. Fig.~\ref{fig:wecare} depicts the scenario presented to survey participants, which was written in a values neutral lens, without explicit mention of any human values. The scenario laid out the objective, context of use, and key requirements. Through the introduction of a common scenario, our aim was to make the survey experience uniform across all the participants who varied in their demographic aspects such as job roles, project experiences, and geographical locations.

\subsubsection{The Decision Points}

After introducing the scenario, we proposed five standard features for the WeCare app. In this research, we identify \textit{standard features} as app functionalities that are common in almost all the apps, without considering any specific scenario. We suggested the following five standard features and asked participants to select the features they wanted to see in the WeCare App.
\begin{itemize}
\item Register- this feature allows the users to provide their necessary information and register with the application.
\item Login - this feature allows the users to provide a correct username/email and password to login to the system.
\item Login (social media) - this feature allows the users to use existing social media to login to the system.
\item Search - this feature allows the users to search within the application. Any settings, information matching with the search string will be the output.
\item FAQ - frequently asked questions are listed and answered.
\end{itemize}
These standard features were suggested upfront in order to save the participants' time spent on coming up with such features while brainstorming in the follow-up sections of the survey. Further, they indirectly acted as example templates that the participants could follow when asked to suggest their own features in the upcoming questions of the survey.

As depicted using the gray diamond shapes in Fig.\ref{fig:surveyflow}, after the standard feature selection, the participants reached the first \textit{decision point (DP1)} of the survey, where the participants decided whether more features are needed to accommodate the requirements mentioned in the scenario other than standard features. If yes, they were given a chance to suggest up to seven new features to the WeCare App. If not, the participants were directed through the alternate route demonstrated using dashed lines in the Survey flow (see Fig.~\ref{fig:surveyflow}).  

When suggesting features, the participants were given a chance to mention the values category they had in their mind. The options list also included the \textit{none of the values} option to indicate the feature was suggested from a values neutral point of view. Then, the participants were presented with a 3-minute video that further explains the importance of having values in software in general. Afterwards, the participants reached the second decision point (DP2), where they were given a chance to change the suggested features or keep them as suggested. We hoped that the video would help participants modify their suggested features to be better aligned with human values. We will discuss the response to the video later in the \textit{Reflections} section \ref{sec:reflections}.

\subsubsection{Values Triggers}
Until this point of the survey, participants suggested features and linked them with human values -- a bottom-up approach. The final section of the survey attempted a top-down approach with \textit{values triggers}. This section showed participants the ten values categories, their definition, and examples as values triggers and asks whether they can identify any features that align with the given human values. This approach allowed them to start with a broader range of values and suggest new features, in addition to the features the participants had suggested earlier. 
All the participants were directed to this section, including those who said `no' in decision point 1. This marked the endpoint of the survey. We will discuss the effect of values triggers in Section~\ref{sec:Findings} and Section~\ref{sec:DiscussionandOurExperience}.  

\subsection{Survey Sampling and Data Collection }

In this survey, we intended to target software practitioners involved in RE-related activities. Therefore, we used a non-probabilistic purposive sampling technique in the study~\cite{baltes2020sampling}. 
We used \textit{Qualtrics} platform to design the survey and, subsequently, we advertised the survey as an anonymous survey link (without any email logging) for RE communities via social media (LinkedIn, Facebook and Twitter) and email lists. The survey attracted nearly 70 practitioners; however, data cleansing resulted in 56 usable responses as we removed responses that did not reach the endpoint in the survey (see Fig.\ref{fig:surveyflow}). Considering the detailed, scenario-based, and partly open-ended nature of the survey, it took approximately 30 minutes 
to complete on an average and generated significant amount of qualitative data to analyse. The effort to attract RE-related participants was successful as 42 of the 56 (75\%) said they were involved in eliciting, analysing, prioritising, or designing software requirements as a part of their job at least a couple of times a week. Another six participants (10.7\%) are involved in RE activities at least couple of time a month while remaining participants mentioned they involved in RE activities couple time a year or very rarely. 
Participants demographics are presented in the Results section \ref{sec:demographics}.

\subsection{Data Analysis} 
The data collected through 56 participants included quantitative and qualitative data; therefore, we used mixed-method analysis to derive the results. Quantitative data mainly emerged from the first section (demographics and values familiarity of participants) of the survey. The quantitative data was analysed using \textit{Qualtrics reports} and Google spreadsheets by the first author of the paper. After the scenario introduction, the survey produced qualitative data, which mainly consisted of the suggested features by the participants. In this survey, we use those suggested features as the primary unit of analysis to understand the effect of explicit consideration of human values in RE activities. 

The data analysis involved three of the authors as the analysts. All of the analysts had a decent understanding of human values and experiences in conducting qualitative analysis.
We applied Socio-Technical Grounded Theory (STGT) \textit{for Data Analysis} \cite{hoda2021STGT} to analyse the qualitative data, using techniques such as open coding, constant comparison, and writing memos. Since the survey responses provided sufficient qualitative data to apply the coding techniques but were not enough (say, as compared to in-depth interview responses) for full theory development, a limited application of STGT \textit{for data analysis} was found suitable \cite{hoda2021STGT}. We selected this approach over other qualitative analysis techniques, such as thematic analysis, because of its (a) rigour that led to multi-dimensional results (presented in section \ref{sec:Findings}) that are \textit{original}, \textit{relevant}, and \textit{dense} as evidenced by the depth of the categories; and (b) reflective practices such as memo writing that led to layered insights and reflections (presented in section \ref{sec:insights} and \ref{sec:reflections}). 



\subsubsection{Open Coding and Feature Categories}

\begin{table*}[]
\centering

\caption{Examples of \textit{STGT for Data Analysis} application \cite{hoda2021STGT} for deriving codes, concepts, and categories from the suggested features.}
\label{tab:featureCategories}
{\renewcommand{\arraystretch}{1.2}
\resizebox{0.98\textwidth}{!}{%
\begin{tabular}{p{0.04\linewidth}p{0.5\linewidth}p{0.13\linewidth}p{0.12\linewidth}p{0.12\linewidth}}

\hline \rowcolor{lightgray} \textbf{Index}&\textbf{Suggested Feature (Raw qualitative data from survey)}& \textbf{Code}& \textbf{Concept}&\textbf{Category}\\\hline


VAL02&``Responsible" & Responsible & Responsible & \multirow{5}{*}{Human Values}\\\cline{1-4}

VAL08&``Ensure users have full control of their experience" & Control experience  & Independent & \\\cline{1-4}

VAL09&``Give some rewarding feelings in application functions" & Reward yourself  & Pleasure & \\\cline{1-4}

VAL16&``Ability to connect via the app based on common
parameters of individuals" & Connect with mutual  & Sense of belonging & \\\midrule


eFR06&``Push notifications should be sent time to time based on the
location of service seeker on nearby service providers." & Location based suggestions & \multirow{4}{*}{\shortstack[l]{Functional\\Requirements}} & \multirow{7}{*}{\shortstack[l]{Requirements\\Type}}\\\cline{1-3}

uFR03&``Ability to recommend the service provider to a friend, Articulate
how the data captured while user sign will be used." & Recommendation services  &  & \\\cline{1-4}

eNR07&``Access to information with less number of clicks/swipes" & Usability  & \multirow{3}{*}{\shortstack[l]{Non-Functional\\Requirements}} & \\\cline{1-3}

uNR01&``Provide physical locations where users can access services at
a kiosk or the like, if they don’t have a phone to use the app" & Accessibility  &  & \\\midrule

eFR12&``Suggestions - notifications for relevant services" & Suggestions
 & \multirow{3}{*}{Epic/Theme level} & \multirow{8}{*}{Granularity}\\\cline{1-3}
eNR08&``Clear and straight forward UI''& UX  &  & \\\cline{1-3}
uFR07&``Portal to connect with each others to build friendship/support without revealing identity" & Create secure interaction &  & \\\cline{1-4}

eFR21&``As a user, I should be able to reserve a service" & Reserve service & \multirow{2}{*}{User story Level} & \\\cline{1-3}

uFR04&``Public to list out items they are willing to donate" & Donation listing & & \\\cline{1-4}

eFR19&``Search should be able to filter by different categories" & Search filters &\multirow{2}{*}{Task Level}  &
\\\cline{1-3}

uFR06&``Providing all possible options under the sex of the person" & Gender options &  & \\\midrule

eFR04&``\textit{Services near me} - allow users to browse local
services" & Search services &\multirow{6}{*}{
Expected feature} & \multirow{11}{*}{\shortstack[l]{Expected\\ Outcome}}\\\cline{1-3}

eFR20&``When the user want to on board on specify service provider,
he/she should be put his credential as a token of responsibility,
thus the provider could have capacity planning beforehand" & Capacity planing &  & \\\cline{1-3}

eNR10&``App should clearly make statement about privacy and which data is being used by the company" & Privacy policy &  & \\\cline{1-4}

uFR04&``Public to list out items they are willing to donate" & Public donation & \multirow{4}{*}{Unexpected Feature} & \\\cline{1-3}

uFR05&``Ability for the homeless to create value through their art/creations (similar to fair trade) facilitate by a platform connected to the apps" & Sell products &  & \\\cline{1-3}

uNR02&``Customize according to the ages" & Personalized UI &  & \\\bottomrule

\end{tabular}
} 
} 
\end{table*}

Open coding was used to identify the \textit{codes} from the suggested features. The suggested features were shared through the open text boxes in the survey and served as the raw qualitative data on which analysis was applied. Using constant comparison, the codes were grouped into \textit{concepts} and concepts into \textit{categories}. An example of the analysis is presented below.

\begin{addmargin}[2em]{2em}

\noindent \textbf{Raw Data:} ``\textit{Push notifications should be sent [from] time to time based on the location of [the] service seeker on nearby service providers}''

\noindent \textbf{Code:} \textit{Location-based suggestions}
\end{addmargin}

\noindent Similarly, other codes such as \textit{recommendation services} were derived from the suggested features. These codes were combined to form a higher-level concept, \textit{functional requirements}.

\begin{addmargin}[2em]{2em}
\noindent \textbf{Concept:} \textit{Functional Requirements}
\end{addmargin}

\noindent Similar concepts were combined to form a category. In this case, the concepts \textit{functional requirements} and \textit{non-functional requirements} were combined to form a higher-level category, \textit{requirements type}.

\begin{addmargin}[2em]{2em}
\noindent \textbf{Category:} \textit{Requirements Type}
\end{addmargin}

Table \ref{tab:featureCategories} presents several such examples of the application of STGT for data analysis to derive the codes, concepts, and categories. Based on the data analysis, four categories were identified: \textit{\textbf{Human Values}}, \textit{\textbf{Requirements Types}}, \textit{\textbf{Granularity}}, and \textbf{\textit{Expected Outcome}} as explained below.

\begin{itemize}
\item \textbf{Human Values as Features [VAL]:}
Deriving the \textit{\textbf{Human Values}} category involved applying open coding and constant comparison as described earlier to extract the value-based codes and then mapping them to the Schwartz model for consistency of terminology at the concept level. For example, participants had suggested individual values (e.g., `helpful' or `forgiving' as features). These features mapped one-to-one with the individual values of the Schwartz model. Further, we found some suggested features were directly derived from human values notion. For example, VAL09--`\textit{give some rewarding feelings in application functions}' or VAL13--`\textit{display the content on the app based on the users traditional values and origin}' are directly linked to Schwartz's values, \textit{pleasure} and \textit{respect for tradition}, respectively. These human value features are captured under the category \textit{\textbf{Human Values}} and are prefixed by \textbf{VAL} for ease of referencing throughout the rest of the paper.

\item \textbf{Requirements Type:}
As described above, through open coding, we identified the category \textbf{\textit{Requirements Type}} to capture the concepts \textit{functional requirements} (e.g., feature eFR04--`\textit{services near me -- allow users to browse local services}') or non-functional requirements (NFRs) (e.g., feature eNR07--`\textit{access to information with less number of clicks/swipes}'). We use prefix \textbf{FR} and \textbf{NR} to identify these classifications respectively. 
     
\item{\textbf{Feature Granularity:}} The third category to be derived from the data analysis was \textbf{\textit{Granularity}}. Since participants were free to suggest features as they liked, without any format constraints or specific guidance, the responses varied in the level of granularity of the features. For example, some of the suggested features were described at the level of implementation details (e.g., feature eFR19-- `\textit{search should be able to filter by different categories}'), which would normally be captured as \textit{tasks} by software teams. On the other hand, some other suggested features were pitched at a more abstract level, without implementation details, otherwise known as \textit{user stories}. Finally, some were described at an even more abstract level (e.g., feature eNR08-- `\textit{clear and straight forward UI}') that could serve as a high-level guidelines, themes, or \textit{epics} and depending on their relative importance, they can be applied as an overarching principle or broken down into specific user stories. This is similar to Bick et al.'s findings and categorisation of agile backlog items to be in a range of coarse-grained (theme or epic) level to fine-grained level (task) \cite{bick2017coordination}.

\item{\textbf{Expected Outcome:}} During the analysis, we observed that features could be categorised based on whether they were suggested using information that was within the scope of the given scenario or by bringing ideas from outside of the scenario (i.e, `\textit{thinking outside-the-box}'). For example, feature eFR06--`\textit{push  notifications should  be  sent [from]  time  to  time  based  on  the location of service seeker on nearby service providers}' is well within the weCare app scenario shared with the participants (see Fig.~\ref{fig:wecare}). However, feature uFR04--`\textit{public to list out items they are willing to donate}' was well outside the details of the weCare app scenario, as the general public was never mentioned as a stakeholder of this application. This led us to classify some suggested features as \textit{expected} (prefix \textbf{e}) and others as \textit{unexpected} (prefix \textbf{u}), from the viewpoint of the scenario exercise. This helped us add an additional layer of detail to our categorisation of functional and non-functional requirements, captured by adding the prefix \textbf{`e'} where the feature was expected and the prefix \textbf{`u'} where it was unexpected, as follows:


\begin{itemize}
    \item \textbf{eFR} - Expected Functional Requirements
    \item \textbf{eNR} - Expected Non-functional Requirements
    \item \textbf{uFR} - Unexpected Functional Requirements
    \item \textbf{uNR} - Unexpected Non-functional Requirements 
\end{itemize}


\end{itemize}

\subsubsection{Direct and Inferred Values Mapping}
In addition to categorizing, we mapped suggested features to Schwartz's values. Almost all the features in the VAL category had a direct mapping for Schwartz's values as they were either value themselves or values related qualities. For other feature categories (eFR, eNR, uFR and uNR), we inferred values based on the feature description and authors' experience in values in software engineering. While inferring, we further identified two different levels based on the easiness to relate a suggested feature to a particular value. Consider feature eFR08-- `\textit{Forum - a place inside the app where users can publish posts and add comments}'. This feature can be easily linked to being \textit{helpful} because a forum and its public posts help each other, i.e. \textit{direct inferred mapping}.
Also, if we take the inferring a step further, it can be seen that a forum may align with being \textit{curious} or looking for \textit{friendship} or \textit{sense of belonging}, i.e. \textit{indirect inferred mapping}.
Altogether, we use three levels of values mappings in this study as follows:
    \begin{itemize}
        \item Mapping level 1 - Direct values mapping (mainly in VAL category)
        \item Mapping level 2 - Direct inferred values mapping
        \item Mapping level 3 - Indirect inferred values mapping
    \end{itemize}

These are denoted by the superscripts x\textsuperscript{1}, x\textsuperscript{2}, and x\textsuperscript{3} respectively on the value names (x) in Tables \ref{VALFeatures}, \ref{EFRFeatures}, and \ref{ENRFeatures}. 

\subsubsection{Role of Memos}
As a part of the STGT process, we wrote `memos' to document the insights generated while performing the open coding activities. While the open coding provided valuable results through categorising the features in terms of human values, requirements types, granularity, and expected outcomes, the memos helped to surface nuanced insights of this study. We draw on these memos in Section~\ref{sec:insights} and ~\ref{sec:DiscussionandOurExperience}, where we share our insights and discussions. Following is an example of a memo created. 

\begin{tcolorbox}[breakable,toprule at break=0mm,bottomrule at break=0mm,title= Memo - \textit{\#ValuesTriggering}]
Probing participants with human values worked, as nearly half of the features were suggested after probing with values in the last question. Is it easier for practitioners to think from values to features (top to bottom) rather than from features to values? Some of the value categories received their first feature just because we probed participants with values!
\end{tcolorbox}

\section{Findings}
\label{sec:Findings}
In this section, we present the findings from the survey analysis. First, we present the outcome of the first section of the survey - \textit{understanding the participant} (see Fig.~\ref{fig:surveyflow}), including participant demographics and their values familiarity. Then we present the outcome of the rest of the survey questions, mainly the feature categorisation, where we present 66 features across five categories -- human values \textbf{(VAL)}, expected functional requirements \textbf{(eFR)}, unexpected functional requirements \textbf{(uFR)}, expected non-functional requirements \textbf{(eNR)}, and unexpected non-functional requirements \textbf{(uNR}). We discuss each of these categories and insights of the respective features. 

\subsection{Participant Demographics}
\label{sec:demographics}
\begin{figure}[htb]
    \centering
    \includegraphics[width=0.5\textwidth]{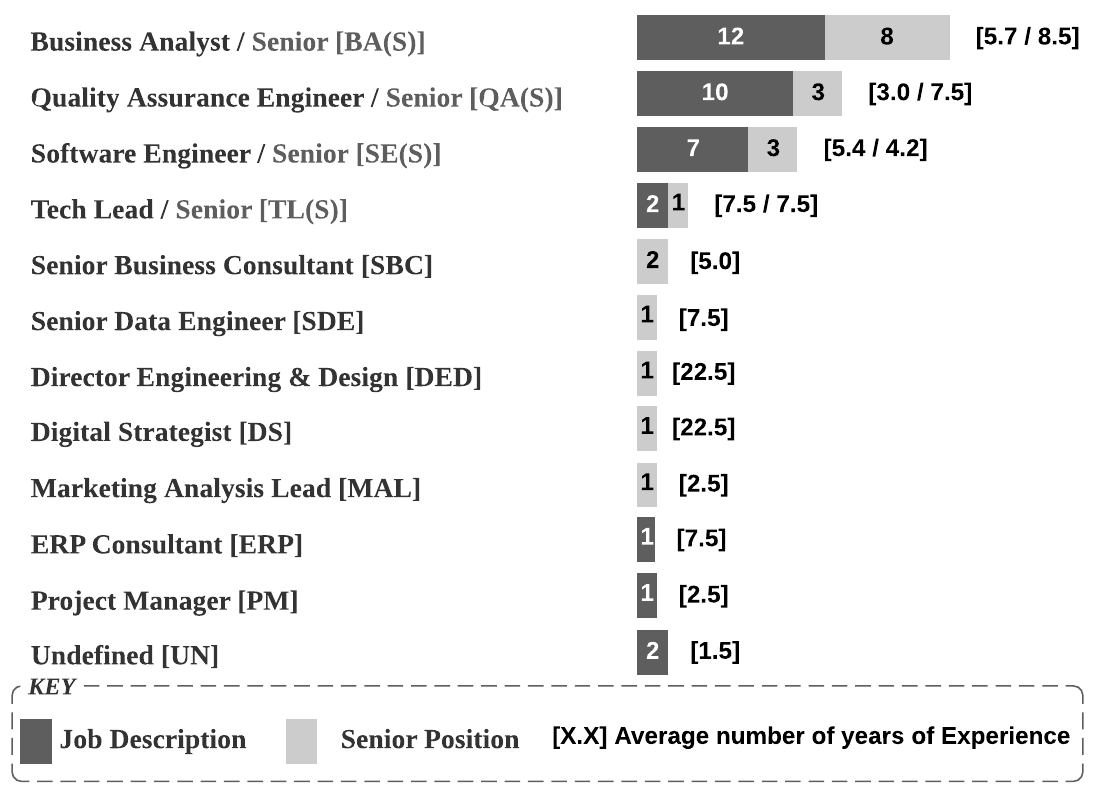}
	\caption{Demographic data of survey participants}
	\label{fig:demographics}
\end{figure}

Out of the 56 participants, 20 (35.7\%) identified as Business Analysts and 13 (23.2\%) as Quality Assurance Engineers, which were the most common job roles among participants. The authors re-categorised some of the similar job roles into commonly known job roles to the best of their knowledge and experience. For example, requirements engineers and business analysts roles were categorised as business analysts. Most participants (26, 46.43\%) had 1-5 years of work experience in the software industry, while 20 (35.71\%) participants had 5-10 years of experience. We also had three participants (5.34\%) with 20 to 25 years of experience in the software industry. We have summarised the demographics of the participants in Fig. \ref{fig:demographics}. To calculate the average years of experience, we use the midpoint of the year category as the \textit{fair} value (for example, if a participant selected 5-10 years as his/her experience, we assumed they had 7.5 years of experience). The overall average of 56 survey participants was calculated as 6.07 years of experience.

\subsection{Values Familiarity} 
\label{subsec:valuesFamiliarity}

\begin{figure}[htb]
    \centering
    \includegraphics[width=0.5\textwidth]{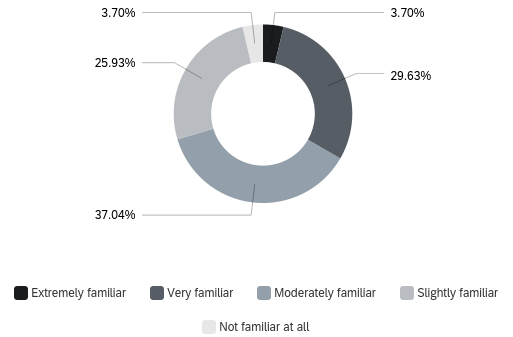}
	\caption{The level of familiarity of values by participants}
	\label{fig:valuesFamiliarity}
\end{figure}

Most participants (70.37\%) were either extremely familiar (3.7\%), very familiar (29.63\%), or moderately familiar (37.04\%) with the values. Fig.~\ref{fig:valuesFamiliarity} shows these levels of familiarity with human values.  


\begin{table*}[]
\centering
\caption{{Suggested \textbf{Values or Values related qualities [VAL]} as features}\break {\vspace{0.5mm}{\scriptsize (\#-suggested by authors; *-suggested by participants; [value]\textsuperscript{X} - x: Values mapping level)}}}
\label{VALFeatures}
\resizebox{\textwidth}{!}{%
{\renewcommand{\arraystretch}{1.3}
\begin{tabular}{p{0.05\linewidth}|p{0.425\linewidth}|p{0.3\linewidth}|p{0.175\linewidth}}

\hline \rowcolor{lightgray}  \textbf{Index}&\textbf{Suggested Features}& \textbf{Individual Value(s)\textsuperscript{\#}}& \textbf{Value Category(s)*} \\ \hline


VAL01&Helpful &Helpful\textsuperscript{1} &Benevolence\\\cline{1-4}

VAL02&Responsible & Responsible\textsuperscript{1} &Benevolence\\\cline{1-4}

VAL03&Forgiving &Forgiving\textsuperscript{1} &Benevolence\\\cline{1-4}

VAL04&Empathy &Helpful\textsuperscript{2},
&
Benevolence\\\cline{1-4}

VAL05&Human connection  & Love\textsuperscript{1}, Sense of belonging\textsuperscript{1} &Benevolence\\\cline{1-4}

VAL06&Independent &Independent\textsuperscript{1} &Self-direction\\\cline{1-4}

VAL07&Situation awareness  &Intelligence\textsuperscript{3} &Self-direction\\\cline{1-4}

VAL08&Ensure users have full control of their experience & Independent\textsuperscript{1}, Freedom\textsuperscript{2} &Self-direction\\\cline{1-4}

VAL09&Give some rewarding feelings in application functions  & Pleasure\textsuperscript{1} &Hedonism\\\cline{1-4}

VAL10&Gender sensitivity  &Self-respect\textsuperscript{1} &Tradition\\\cline{1-4}

VAL11&Social considerations  & Respect for tradition\textsuperscript{1}  &Tradition\\\cline{1-4}

VAL12&The app should not asking private data that is not adhere with tradition / data that is not related to the provided service. For example, asking "religion" for "food" can be changed with providing the food menu, so the user could choose which service provider that provide food that adhere to his/her religion & Respect for tradition\textsuperscript{1} &Tradition\\\cline{1-4}

VAL13&Display the  content on the app based on the users traditional values and origin &Respect for tradition\textsuperscript{1}
&Tradition\\\cline{1-4}

VAL14&Emotional situation & Sense of belonging\textsuperscript{2}  &Security\\\cline{1-4}

VAL15&Join hands - to help others in need around you &Helpful\textsuperscript{1}, Social power\textsuperscript{2}, Sense of belonging\textsuperscript{2} &Security\\\cline{1-4}

VAL16&Ability to connect via the app based on common parameters of individuals &Sense of belonging\textsuperscript{1}, Helpful\textsuperscript{3}&Security\\\cline{1-4}

VAL17&Key success indicators - a personalised metric created and set by the individual service seeker in a checklist, that they can tick off to enable self-approval or self-worth. & Successful\textsuperscript{1} &Achievement\\\bottomrule

\end{tabular}
} 
} 
\end{table*}

The follow-up question revealed the value categories that participants often consider when they develop software in general. Participants were allowed to select multiple values categories. The percentage of familiarity for each value category is presented in Fig.~\ref{fig:valueswithfeatures}. 
Unsurprisingly, \textit{Security} -- the well know software quality aspect -- scored the highest popularity (62\%) while \textit{Hedonism} recorded as the least popular (18\%) values category. We discuss the way this value category popularity may have affected the suggested features under \textit{\#FamiliarityImpact} in the \textit{Insights} section \ref{sec:insights}.  


\subsection{Standard Features Selection}
Next, we discuss the results of the survey after the introduction of the scenario. The first task was to select the standard features from a given list. 
Registration, Search, and FAQs standard features were selected by more than 75\% of the participants, while Login with or without social media recorded less popularity (around 65\%). This outcome indicates that the participants thought about the \textit{privacy} of users, i.e., homeless people. To this end, we found suggested features such as eFR03 (see Table~\ref{EFRFeatures}) and eNR13 (see Table~\ref{ENRFeatures}) also suggested as being anonymous within the WeCare platform. 


\subsection{Human Values as Features}

We identified 17 values or values related qualities [VAL01 -- VAL17] suggested by the participants (see Table~\ref{VALFeatures}). These suggestions demonstrate that practitioners are capable of identifying values that are aligned with a given scenario. A feature like VAL12 -- `\textit{The app should not asking private data that is not adhere with tradition ... user  could  choose  which service  provider  that  provide  food  that  adhere  to  his/her religion}' shows that the participant made a clear link between the requirement and the value, \textit{Tradition}. Further, we identified evidence that suggests that explicit thinking about human values can alter a typical software feature to better align with values requirements. For example, a standard chat function would enable users to connect with other users; however, VAL16 (see Table~\ref{VALFeatures}) suggests connecting people based on `\textit{common parameters of individuals}', i.e., mutual interest. This example demonstrates that values thinking during RE activities would give an extra dimension to typical software features by adding a purpose, i.e., answering \textit{why?} someone wants to develop a particular feature in the first place. 
However, some of the suggested features were very short in description and used the same terms as the values themselves (e.g., VAL01--`\textit{Helpful}', VAL02--`\textit{Responsible}', VAL03--`\textit{Forgiving}'). It is clear that such short descriptions and listing of value names does not assist with operationalizing such features in practicality. This aligns with one of our previous findings~\cite{hussain2020human}, the inability to translate human values into features being one of the common challenges in operationalizing human values in SE. It also highlights the importance of identifying the granularity level of features that are brainstormed at the early stages of RE. We further discuss this idea in Section~\ref{findings:granularity}. 


\begin{table*}[]
\centering
\caption{Suggested \textbf{Expected Functional Requirements [eFR]} \break \vspace{0.5mm} {\scriptsize (\#-suggested by authors; *-suggested by participants; [value]\textsuperscript{X} - x: Values mapping level)}}
\label{EFRFeatures}
\resizebox{\textwidth}{!}{%

\renewcommand{\arraystretch}{1.3}
\begin{tabular}{p{0.05\linewidth}|p{0.425\linewidth}|p{0.3\linewidth}|p{0.15\linewidth}}

\hline \rowcolor{lightgray}  \textbf{Index}&\textbf{Suggested Features}& \textbf{Individual Value(s)\textsuperscript{\#}}& \textbf{Value Category(s)*} \\ \hline
\multicolumn{4}{c}{\textit{Login and Registration}}\\\cline{1-4}

eFR01&Different login options for service seekers and service providers &Security\textsuperscript{2}, Independence\textsuperscript{3} &Security\\\cline{1-4}

eFR02&Need for the homeless to be registered with the state; entertainment apps/centers to allow access at reduced/zero rates with authentication via app &Security\textsuperscript{2}, Wealth\textsuperscript{2}, Pleasure\textsuperscript{2} &Hedonism\\\cline{1-4}

eFR03&Guest access, certain feature (like Search) should be able to be used without giving credential (register/login). The app should be able to create ephemeral userID (cookies) that identify this user. Hence, subsequent access to the server from same app/device/userId can be treated as same "user" & Privacy\textsuperscript{2}, Independent\textsuperscript{2} &Self-direction\\\cline{1-4}

\multicolumn{4}{c}{\textit{Location based features}}\\\cline{1-4}
eFR04&"Services near me" - allow users to browse local services & Helpful\textsuperscript{2}, Capable\textsuperscript{3}& Benevolence\\\cline{1-4}

eFR05&Google map or any other map integration to give directions to service seekers to find service providers location &Helpful\textsuperscript{2}&Benevolence\\\cline{1-4}

eFR06&Push notifications should be sent time to time based on the location of service seeker on nearby service providers. &Helpful\textsuperscript{3} &Benevolence\\\cline{1-4}


eFR07&Map service to show the nearest service provider /service seeker's locations  &Helpful\textsuperscript{3} &Benevolence\\\cline{1-4}

\multicolumn{4}{c}{\textit{Help Desk and Forum}}\\\cline{1-4}
eFR08&Forum - a place inside the app where users can publish posts and add comments&Helpful\textsuperscript{2}, Curious\textsuperscript{3}, Friendship\textsuperscript{3}, Sense of Belonging\textsuperscript{3} &Benevolence\\\cline{1-4}

eFR09&Posting queries/help wanted - will allow the service seeker to personalise their need and request for assistance for themselves &Curious\textsuperscript{2}, Helpful\textsuperscript{2}, Freedom\textsuperscript{2}, Independence\textsuperscript{2} &Self-direction\\\cline{1-4}

eFR10&Chat service / hot-line & Curious\textsuperscript{2}, Helpful\textsuperscript{2}, &Achievement\\\cline{1-4}

eFR11&Support - get support for contacting services from a help desk, should be accessible throughout the app &Helpful\textsuperscript{2},  Sense of belonging\textsuperscript{3} &Security\\\cline{1-4}

eFR12&Suggestions - notifications for relevant services &Helpful\textsuperscript{2} &Benevolence\\\cline{1-4}

\multicolumn{4}{c}{\textit{Rating and Feedback}}\\\cline{1-4}
eFR13&Rating systems for eligibility criteria for service seekers and service providers  &Helpful\textsuperscript{3} &Achievement\\\cline{1-4}

eFR14&Reviews/Rating system for service providers. So future service seekers will be able to get an understanding on  how responsible service providers are &Helpful\textsuperscript{2}, Curious\textsuperscript{3}, Security\textsuperscript{2}, Healthy\textsuperscript{2} &Benevolence\\\cline{1-4}

eFR15&Feature should be able to show the services based on the user reviews and ratings &Helpful\textsuperscript{3}, Social recognition\textsuperscript{3}, Freedom\textsuperscript{3} &Security\\\cline{1-4}

eFR16&Ways to leave the feedback \& recognise providers based on the user experience &Helpful\textsuperscript{3}, Social recognition\textsuperscript{2} &Power\\\cline{1-4}

eFR17&The public who are part of the service system to be tiered based on their contribution and thereafter recognized (similar to how contributors to google maps/local guides are treated today) & Social recognition\textsuperscript{2}, Successful\textsuperscript{2}, Influential\textsuperscript{2} &Power\\\cline{1-4}

\multicolumn{4}{c}{\textit{Search related features}}\\\cline{1-4}
eFR18&Feature should be able to save their search criteria&Helpful\textsuperscript{2}, Freedom\textsuperscript{3}, Security\textsuperscript{3}, Capable\textsuperscript{3} &Benevolence\\\cline{1-4}

eFR19&Search should be able to filter by different categories & Capability\textsuperscript{2}, Helpful\textsuperscript{3} &Self-direction\\\cline{1-4}

\multicolumn{4}{c}{\textit{Other features}}\\\cline{1-4}

eFR20&When the user want to on board on specify service provider, he/she should be put his credential as a token of responsibility, thus the provider could have capacity planning beforehand & Capable\textsuperscript{3}, Helpful\textsuperscript{2},  Security\textsuperscript{3} &Benevolence\\\cline{1-4}

eFR21&As a user, I should be able to reserve a service & Independent\textsuperscript{2}, Capable\textsuperscript{2} &Achievement\\\cline{1-4}

eFR22&Favourites - this feature will allow the service seekers to list the type of services they are interested in from a list of existing services, so that they can be marked as favourites.&Helpful\textsuperscript{2}, Authority\textsuperscript{3}, Freedom\textsuperscript{3}, Independent\textsuperscript{3} &Benevolence\\\cline{1-4}

eFR23&Dashboard with icons showing access to services &Helpful\textsuperscript{3} &Self-direction\\\cline{1-4}

eFR24&Ability to select the details and content to be displayed &Freedom\textsuperscript{2}, Independence\textsuperscript{2} &Self-direction\\\bottomrule

\end{tabular}
}
\end{table*}

\begin{table*}[]
\centering
\caption{Suggested \textbf{Expected Non Functional Requirements [eNR]} with related NFRs \break \vspace{0.5mm} {\scriptsize (\#-suggested by authors; *-suggested by participants; [value]\textsuperscript{X} - x: Values mapping level)}}
\label{ENRFeatures}
\resizebox{\textwidth}{!}{%

\renewcommand{\arraystretch}{1.3}
\begin{tabular}{p{0.05\linewidth}|p{0.325\linewidth}|p{0.2\linewidth}|p{0.2\linewidth}|p{0.15\linewidth}}

\hline \rowcolor{lightgray}  \textbf{Index}&\textbf{Suggested Features}&\textbf{NFR(s)\textsuperscript{\#}}& \textbf{Individual Value(s)\textsuperscript{\#}}& \textbf{Value Category(s)*} \\ \hline

eNR01&Ability to use it on another persons phone &Portability& Helpful\textsuperscript{2} &Benevolence\\\cline{1-5}

eNR02&Ability to access it via a browser& Portability, Accessibility &Helpful\textsuperscript{2}, Independent\textsuperscript{3}  &Benevolence\\\cline{1-5}

eNR03&Accessibility settings&Accessibility &Universalism\textsuperscript{1} &Universalism\\\cline{1-5}

eNR04&View the content of the app in user preferred language &Accessibility& Helpful\textsuperscript{2}, Equality\textsuperscript{3}, 
Helpful\textsuperscript{2} &Universalism\\\cline{1-5}

eNR05&Home page with easy access icons for most popular/important services &Accessibility, Usability&Helpful\textsuperscript{2} &Benevolence\\\cline{1-5}

eNR06&Easy navigation &Usability& Helpful\textsuperscript{2} &Self-direction\\\cline{1-5}

eNR07&Access to information with less number of clicks/swipes &Usability&Curious\textsuperscript{3}, Helpful\textsuperscript{2} &Self-direction\\\cline{1-5}

eNR08&Clear and straight forward UI &Usability& Helpful\textsuperscript{3} &Self-direction\\\cline{1-5}

eNR09&Consent settings - allow users to update their consent for location tracking and other privacy considerations. Automatically runs as part of registration, can be updated at any time later& Privacy & Privacy\textsuperscript{1}, Freedom\textsuperscript{2} &Self-direction\\\cline{1-5}

eNR10&App should clearly make statement about privacy and which data is being used by the company & Privacy& Privacy\textsuperscript{1}, Loyal\textsuperscript{2} &Security\\\cline{1-5}


eNR11&Remove data from server. Anytime, the app should have capability to remove traces/data from the local/server&Privacy & Capability\textsuperscript{1}, Privacy\textsuperscript{3} &Self-direction\\\cline{1-5}





eNR12&To link their financial information via Open Banking initiative  &Privacy &Privacy\textsuperscript{2}, Helpful\textsuperscript{3}, Security\textsuperscript{3} &Stimulation\\\cline{1-5}









eNR13&Ability to register without an email address or proof of identity &Privacy, Security &Security\textsuperscript{2}, Privacy\textsuperscript{2} &Security\\\cline{1-5}


eNR14&A feature to add the live location of the service seeker with his consent &Privacy, Security &Security\textsuperscript{2}, Privacy\textsuperscript{2}, Helpful\textsuperscript{2} &Security\\\cline{1-5}

eNR15&Rules and regulations for service seekers and service providers  &Security& Social order\textsuperscript{2}, Obedient\textsuperscript{2} &Security\\\bottomrule






\end{tabular}
}
\end{table*}

\begin{table*}[]
\centering

\caption{Suggested \textbf{Unexpected Functional Requirements [uFR]} and \textbf{Unexpected Non Functional Requirements [uNR]} \break \vspace{0.5mm} {\scriptsize(\#-suggested by authors; *-suggested by participants; [value]\textsuperscript{X} - x: Values mapping level)}}

\label{UFRUNRFeatures}
\resizebox{\textwidth}{!}{%

\renewcommand{\arraystretch}{1.3}
\begin{tabular}{p{0.05\linewidth}|p{0.425\linewidth}|p{0.3\linewidth}|p{0.175\linewidth}}

\hline \rowcolor{lightgray}  \textbf{Index}&\textbf{Suggested Features}& \textbf{Individual Value(s)\textsuperscript{\#}}& \textbf{Value Category(s)*} \\ \hline

\multicolumn{4}{c}{\textit{\textbf{Unexpected Functional Requirements [uFR]}}}\\\cline{1-4}

uFR01&Option to flag 'in distress' or 'help required' to indicate their current status  &Helpful\textsuperscript{2}, Love\textsuperscript{2}, Security\textsuperscript{2} &Benevolence\\\cline{1-4}

uFR02&Allow service seekers to request to be contacted, rather than having to search &Helpful\textsuperscript{2}, Humble\textsuperscript{2} &Benevolence\\\cline{1-4}

uFR03&Ability to recommend the service  provider to a friend, Articulate how the data captured while user sign will be used. &Social\textsuperscript{2}, Power\textsuperscript{2}, Privacy\textsuperscript{2}&Benevolence\\\cline{1-4}

uFR04&Public to list out items they are willing to donate &Helpful\textsuperscript{3}, Capable\textsuperscript{3} &Benevolence\\\cline{1-4}

uFR05&Ability for the homeless to create value through their art/creations (similar to fair trade) facilitate by a platform connected to the apps & Freedom\textsuperscript{2}, Independence\textsuperscript{3}, Creativity\textsuperscript{3} &Self-direction\\\cline{1-4}

uFR06&Providing all possible options under the sex of the person  & Equality\textsuperscript{2}, Social justice\textsuperscript{2}, Humble\textsuperscript{2}, Politeness\textsuperscript{2} , Social recognition\textsuperscript{2} &Universalism\\\cline{1-4}

uFR07&Portal to connect with each others to build friendship/support without revealing identity & Friendship\textsuperscript{1}, Privacy\textsuperscript{2}, Helpful\textsuperscript{2}, Security\textsuperscript{2}, Sense of belonging\textsuperscript{2} &Benevolence\\\cline{1-4}


\multicolumn{4}{c}{\textit{\textbf{Unexpected Non Functional Requirements [uNR]}}}\\\cline{1-4}

uNR01&Provide physical locations where users can access services at a kiosk or the like, if they don't have a phone to use the app - \textit{Accessibility\textsuperscript{\#}} &Helpful\textsuperscript{2}, Capable\textsuperscript{2} &Universalism\\\cline{1-4}

uNR02&Customize according to the ages - \textit{Accessibility\textsuperscript{\#}}  & Respect for tradition\textsuperscript{2}, Universalism\textsuperscript{3}&Conformity\\\cline{1-4}

uNR03&The user should have choice to put/remove their data on the server. Hence "Remove my data" as a feature is mandatory - \textit{Privacy\textsuperscript{\#}} &Privacy\textsuperscript{2}, Independent\textsuperscript{2}, Freedom\textsuperscript{2} &Self-direction\\\bottomrule

\end{tabular}
}
\end{table*}

\subsection{Requirements Type} 
Through the STGT data analysis, detailed in section \ref{sec:Methodology}, 
out of the 66 suggested features, 31 were identified as functional requirements (eFR01 to eFR24 (Table~\ref{EFRFeatures}) and uFR01 to uFR07 (Table~\ref{UFRUNRFeatures})), while 18 features categorised as non-functional requirements (eNR01 to eNR15 (Table~\ref{ENRFeatures}) and uNR01 to uNR03 (Table~\ref{UFRUNRFeatures})). 

The functional requirements in Table~\ref{EFRFeatures} mainly addressed the scenario requirements. We have identified different sub-collections within eFR features aligning with major functional components of the WeCare app such as login and registration, location-based feature, rating \& feedback (see Table~\ref{EFRFeatures}).
We found participants have suggested similar functional components; however, the descriptions of the functions have given different dimensions to the feature. For example, feature eFR10 suggests having `\textit{chat service / hot-line}' while feature eFR11 suggests the same with a purpose as `\textit{support - get support for contacting services from a help desk, should be accessible throughout the app}' (see Table~\ref{EFRFeatures}). Also. The latter expresses the need for accessibility, a quality aspect of the feature in addition to what it should do.

Moreover, the participants were found to draw ideas from real-world applications to suggest some features for the WeCare app scenario. For instance, feature eFR17 proposes `\textit{the public who are part of the service system to be tiered based on their contribution and thereafter recognized (similar to how contributors to google maps/local guides are treated today)}'. Such features (e.g., eFR03, eFR05, eFR17, eNR12, uFR05)
indicate that if given a scenario, practitioners would be able to draw ideas from similar experiences and contexts, which supports scenario-based thinking as potentially an effective tool toward operationalizing human values in RE.

Suggested non-functional requirement features mainly addressed quality aspects such as portability, accessibility, usability, and privacy. Under values mapping activity, we have mapped these features to Schwartz's values as demonstrated in Table~\ref{ENRFeatures} and Table~\ref{UFRUNRFeatures}. Most of these mappings were in level 2 or 3, i.e., inferred values mapping with direct or indirect links. On a related note, human values are often confused with non-functional requirements. For example, eNR09 to eNR15 (see Table~\ref{ENRFeatures}) mainly address the privacy of users, which could be categorised as an NFR or a value. However, the feature description were closer to the quality aspects of the app; therefore, we categorised them as non-functional requirements.

Similar to functional requirements, some features were suggested with their purpose. For example, feature uNR03 (see Table~\ref{UFRUNRFeatures}) suggested to develop a feature called \textit{`remove my data'}  because `\textit{user should have the choice to put/remove their data on the server}'.
Such outcomes suggest that in both functional and non-functional requirements, values thinking would encourage practitioners to think about the purpose of the feature they develop.  

A part of this scenario-based survey activity was to map the suggested features to human values categories in Schwartz's theory of basic human values~\cite{Schwartz1992}. Participants mapped all (66) suggested features to values categories. 
All the features are visualised under their participant classified value categories, as depicted by superimposing on the Schwartz model's circular structure (see Fig.~\ref{fig:valueswithfeatures}). 

\subsection{Feature Granularity}
\label{findings:granularity}

\begin{figure*}[htb]
    \centering
    \includegraphics[width=\textwidth]{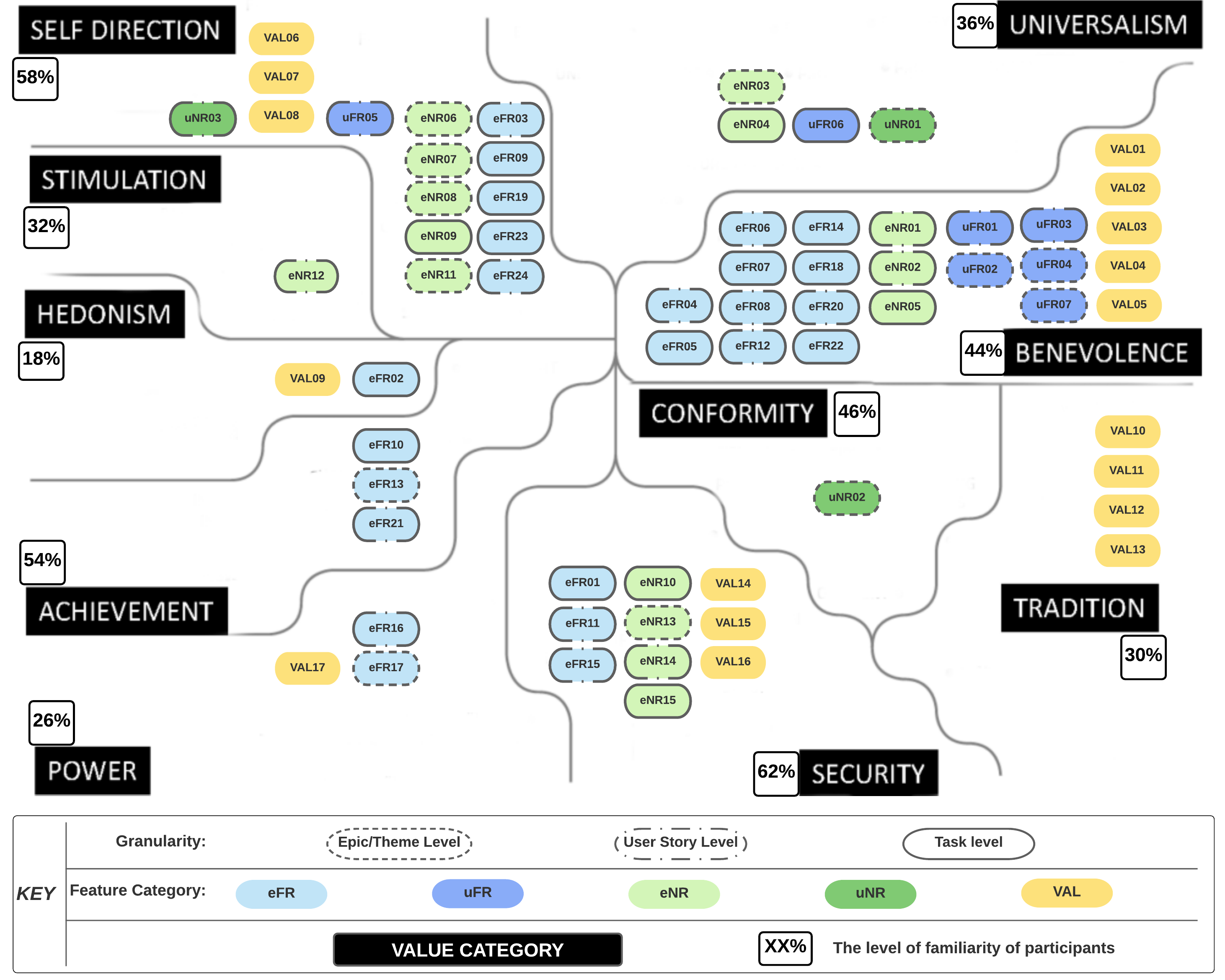}
	\caption{Schwartz's theory (circular structure)~\cite{schwartz1992universals} (adapted from~\cite{holmes_blackmore_hawkins_wakeford_2011}). Words in black boxes are values categories. All the suggested features are superimposed over Schwartz's values categories}
	\label{fig:valueswithfeatures}
\end{figure*}
We identified three granularity levels: \textit{epic/theme} level, \textit{user story} level, and \textit{task} level at which the suggested features were pitched. We have demonstrated the feature granularity in Fig.~\ref{fig:valueswithfeatures} using the different outlines for feature bubbles. For the features in the VAL category (yellow bubbles), we did not define any granularity level as almost all of them are still at a highly abstract level. 
We found that most of the Expected Functional Requirements (eFR) (light blue bubbles in Fig.~\ref{fig:valueswithfeatures}) were suggested at either the user story level or the task level. Practitioners work with such functional requirements every day, and perhaps that allowed the participants to produce more fine-grained level features~\cite{bick2017coordination}. For example, though we did not request participants to use any particular format when suggesting features, one of the participants suggested a feature using the user story format -- `\textit{as a user, I should be able to reserve a service}' (eFR21).

Expected non-functional requirements (eNR) (light green bubbles) and unexpected functional requirements (uFR) (dark blue bubbles) had a mixed result in terms of granularity between epic/theme level and user story level. It was noted that the non-functional and unexpected features were suggested at more coarse-grained levels, which aligns with similar findings in agile planning contexts ~\cite{bick2017coordination}. For example, the three unexpected non-functional requirements (uNR) items (dark green bubbles) support this argument as two of them were pitched at the epic level, while the other on the user story level.   

Converting abstract concepts such as human values into actionable software tasks is one of the critical challenges in operationalizing human values in SE~\cite{hussain2020human}. These findings suggest that considering human values while conducting RE activities may reveal features with different granularity levels. Therefore, being conscious about the level of granularity of the features in software design may help operationalize human values in RE. For instance, while conducting RE activities, a practitioner may explicitly label the granularity of the design choices and try to brainstorm less abstract features and more towards well-defined task levels. To give an example, one participant suggested feature eNR06-- `\textit{Easy navigation}', which is pitched at the epic/theme level, while another participant suggested eNR05-- `\textit{Home page with easy access icons for most popular/important services}', which is a more actionable task level feature. 


\subsection{Expected Outcome}
We found a total of 39 \textit{expected} features, both functional (eFR01 to eFR24) and non-functional (eNR01 to eNR15) and 10 \textit{unexpected} features (uFR01 to uFR07 and uNR01 to uNR03). As explained in section \ref{sec:Methodology}, our definition of expected features is bound to whether the feature emerged directly from the given information in the scenario (expected) or whether the participants drew on ideas outside the scenario and suggested the feature (unexpected).

We found some of the feature descriptions among expected features that depict that the scenario-based approach eased the participants' thinking process and helped them  develop the purpose of the feature they proposed, i.e., \textit{why} you need this feature? For example, eFR14 said `\textit{reviews/rating system for service providers.  So future service seekers will be able to get an understanding of how responsible service providers are}'. With the use of the `\textit{so}' the participant went on to describe the rationale for \textit{why} the feature is required. In this case, by providing a feature to review the quality of the service, it seems they want to ensure service providers are \textit{responsible}, thereby focusing on the service seekers' \textit{quality of life}. In feature eFR20, a participant suggested that `\textit{When the user wants to on board on specify [specific] service provider, he/she should be put his credential as a token of responsibility; thus the provider could have capacity planning before-hand}'. These examples, in particular, demonstrate that participants had thought about the different stakeholders involved and had considered the purpose of the feature from service seeker's and service provider's points of view. This leads to an interesting discussion on \textit{values trade-offs}~\cite{friedman2002value}, where one stakeholder's values might positively or negatively affect the (rest of the) values of the same stakeholder or another stakeholder.

We found unexpected features as interesting ideas that demonstrated creativity and the ability of the participants to draw ideas from potentially their professional software engineering experience and/or personal worldviews as software users. For example, the feature uNR01-- `\textit{provide physical locations where users can access services at a kiosk or the like, if they don't have a phone to use the app}' not only goes beyond the scenario information but challenges the  scenario assumption that the homeless have access to mobile phones. The participant proposes to make everyone \textit{capable} of using the WeCare app with or without mobile, a precise alignment with value category \textit{Universalism}.
Further, in the feature uFR05, a participant suggests using the WeCare platform as a source of income by `\textit{create value through their [users'] art/creations  (similar to fair trade) facilitate by a platform connected to the apps}'. uFR04 has extended the stakeholder list of the scenario by suggesting the public `\textit{to list out items they are willing to donate}'. 

These examples suggest that explicit consideration of human values can extend the thinking boundaries of practitioners and enable them to come up with more features, feature options, identify more stakeholders, and their roles that make the software design better aligned with human values.


\section{Insights and Reflections} 

\subsection{Insights}
\label{sec:insights}

Based on careful consideration of the results and drawing on our memos, we present some key insights.\\


\noindent \textbf{\textit{\#FamiliarityImpact}}: Based on the evidence, we were able to draw an insight about the potential impact of values familiarity on values elicitation. In subsection~\ref{subsec:valuesFamiliarity}, we have discussed the values familiarity of participants (presented as percentages in greyed outlined boxes in Fig.~\ref{fig:valueswithfeatures}, next to the value category name). The categories with relatively higher familiarity levels visibly have more suggested features. For example,  \textit{Benevolence} (familiarity 44\%), \textit{Self Direction} (58\%) and \textit{Security} (62\%) have absorbed 23, 15 and 10 features, respectively, making them the top three value categories with the most number of features. Similarly, less familiar categories like \textit{Hedonism} and \textit{Power} have lesser number of suggested features (see Fig.~\ref{fig:valueswithfeatures}). This pattern suggests that more the practitioners are aware and familiar with particular values, the more features they are likely to derive aligned with such values. However, \textit{Conformity} is an exception in this regard, where nearly half the participants (46\%) acknowledged their familiarity with this value category but only one feature was suggested in this category (uNR02).\\

\noindent \textit{\textbf{\#ValuesTriggering:}}
We tried both: \textit{feature-driven value mapping}, brainstorm features, then map to values and \textit{values-driven feature mapping}, triggering with values to brainstorm features. At a first glance, it seems both approaches are almost equally effective, as 29 out of 66 features were suggested at the last ``values triggers'' question of the survey (see Fig.~\ref{fig:surveyflow}). However, the fact that the participants were able to collectively derive 29 \textit{additional} features, on top of the ones they had already identified, as a direct result of values triggering can be seen as a significant impact of the values-driven feature mapping approach, and the process of applying both approaches. In addition to its impact on deriving more features, the value triggers were also seen to impact \textit{outside-the-box} thinking and the elicitation of unexpected features. A majority of the \textit{unexpected} features (9 out of 10) were suggested as a response to the values triggering question. More surprisingly, 6 of these were suggested by participants who had initially responded that they did not have more features to add earlier on in the survey (at decision point DP1, see Fig. \ref{fig:surveyflow}) but went on to identify more features using the values-driven feature mapping approach.\\


\noindent \textit{\textbf{\#ValuesConflict:}} Using a circular structure to his model, Schwartz explains interlinks between values. Values located closer to each other are complementary, whereas values further apart tend to be in tension with each other~\cite{Schwartz1992}. Following this principle, it can be suggested that the suggested features listed in a category such as  \textit{Benevolence} may complement the features in the \textit{Universalism} category, while features listed under \textit{Self-direction} may be in conflict with implementing feature listed in \textit{Security}, and vice-versa. For example, the feature VAL13, in the \textit{Tradition} value category suggests `\textit{display the content on the app based on the user's traditional values and origin}', while the feature eFR03 from \textit{Self-direction} category --on opposite end of the model-- requests to implement `\textit{guest access}' which can be an obstacle to collecting the data, such as user's traditional values and origin, necessary to implement the feature VAL13. This leads us to our next insight about values trade-off.\\

\noindent \textit{\textbf{\#ValuesTrade-offs:}} We identified that values trade-offs could occur for the same stakeholder (as described above) or between different stakeholders. For example, eFR10 and eFR11 (see Tabel~\ref{EFRFeatures}) both suggest having chat services/hot-line or a help desk. These features are \textit{helpful} for service seekers; however, such external services might negatively affect the \textit{wealth} of the government, i.e., cost for recruiting people, conducting training and maintaining the help desk. 
Though it was not our intention to handle values trade-offs, including prioritisation, in this study, we acknowledge the importance of handling them and will continue our future research to resolve them.\\ 




\subsection{Reflections}
\label{sec:reflections}
Conducting this scenario-based survey study was an interesting experience. We share some reflections which may be helpful for other researchers and practitioners.\\

\noindent \textit{\textbf{\#ExperienceHelps:}} We further analysed the demographics of practitioners who proposed the unexpected features. Practitioners who proposed unexpected features had an average of 9 years of experience, while the same for the entire sample was 6.07 years. Moreover, six out of ten features were suggested by practitioners who held senior positions, mainly the BA roles, suggesting that maturity in the software industry allows practitioners to consider a wide range of issues associated with software requirements and design, to  \textit{think-outside-the-box} where the `\textit{box}' is the given scenario, and to elicit specific values. \\

\noindent \textit{\textbf{\#ChoiceofScenario:}} The choice of the scenario is likely to have an impact on the values categories elicited from the scenario-based survey. In this study, the scenario used was based on a proposed mobile app for homeless people. While the scenario itself was written with a values neutral lens, the nature of the application domain, i.e. providing shelter for the homeless, is likely to have elicited certain categories of values over others. For example, \textit{Benevolence} and \textit{Self-direction} were the value categories which elicited a maximum number of value items based on self-identified by the participants and as inferred through the analysis of the features. These categories align with the values perceived to be demanded by the users (e.g. helpful, responsible) and those seen to be supported by the app (e.g. freedom, independence) respectively. It may also suggest why only one feature was suggested in the \textit{Conformity} category, despite nearly half the participants being familiar with this value category.\\

\noindent \textit{\textbf{\#ResearchAdaptations:}} We had originally planned to conduct the study as in-person workshops. However, the global Covid-19 pandemic imposed strict and extended lockdowns in Melbourne, as in many parts of the world, forcing us to consider other ways of continuing our research study. The research team brainstormed alternatives. Considering zoom fatigue and the need for schedule flexibility for better work-life balance while working from home during such challenging times, we decided to proceed with a survey which could be filled asynchronously in the participant's own time. We played with the idea of embedding the scenario in the survey. At first, this seemed difficult but after several rounds of reviews, we managed to design a reasonable survey flow. We were pleasantly surprised by the number of participants and their sincere working through the survey responses, spending 30 minutes on an average and eliciting 66 features all together. Our experience suggests a well-designed scenario-based survey is a reasonable data collection technique, especially when under physical and time constraints.\\

\noindent \textit{\textbf{\#ValuesVideo:}} We had hoped that a visual introduction of values through a 3-minute video (see Fig.~\ref{fig:surveyflow})  would  help  participants modify  their  suggested features (in the step after the video)  to  be  better  aligned  with  human  values. However, the feature modification step was highly unpopular as almost all the participants marked suggested features as either `\textit{values are already considered}' or `\textit{keep feature as it is}'. The level of values familiarity of participants in this study, as depicted in Fig.~\ref{fig:valuesFamiliarity}, suggests a majority of participants (70.37\%) were well-positioned to extract values. Therefore, it is likely that they were confident about the initial feature suggestions. However, such video introduction might be helpful with a different cohort of participants with lesser values awareness. 



\section{Discussion and Implications} 
\label{sec:DiscussionandOurExperience}



We conducted a scenario-based survey research study to address the research question, \textbf{\textit{what is the impact of considering human values explicitly in the early requirements engineering activities?}} 
In response to the RQ, the results show that considering human values explicitly while conducting requirements analysis registers several impacts. It helps practitioners to:

\begin{itemize}
    \item identify human values that are applicable to a given scenario (VAL), 
    \item associate purpose with the features they develop in their day-to-day life (eFR, eNR), considering the important \textit{why} question, instead of \textit{jumping into} software development,
    \item think \textit{outside-the-box}, beyond the given scenario, and draw ideas from their life experiences (uFR, uNR), and 
    \item build connections between software features and human values (eFR, eNR, uFR, uNR).
\end{itemize}

Overall, the explicit consideration of human values in the early RE activities has a strong potential to enable practitioners to concretely identify and align human values with software requirements -- previously identified as a key challenge of operationalizing humans in software engineering~\cite{Mougouei2018}. Furthermore, we argue that explicit consideration is valuable and essential to developing software that demonstrates and respects human values. Such explicit consideration is likely to lend purpose to SE practitioners while developing software as they can clearly identify and make connections between the requirements they are fulfilling and stakeholders' values. Given the success of the scenario-based survey, we also suggest that scenario-based thinking as a good approach to implementing the connection between features and values. \\

\noindent \textit{\textbf{Practical Implications -- Scenario-Based Values Elicitation}}
\label{descussion.fourstep}

\begin{figure}[htb]
    \centering
    \includegraphics[width=0.39\textwidth]{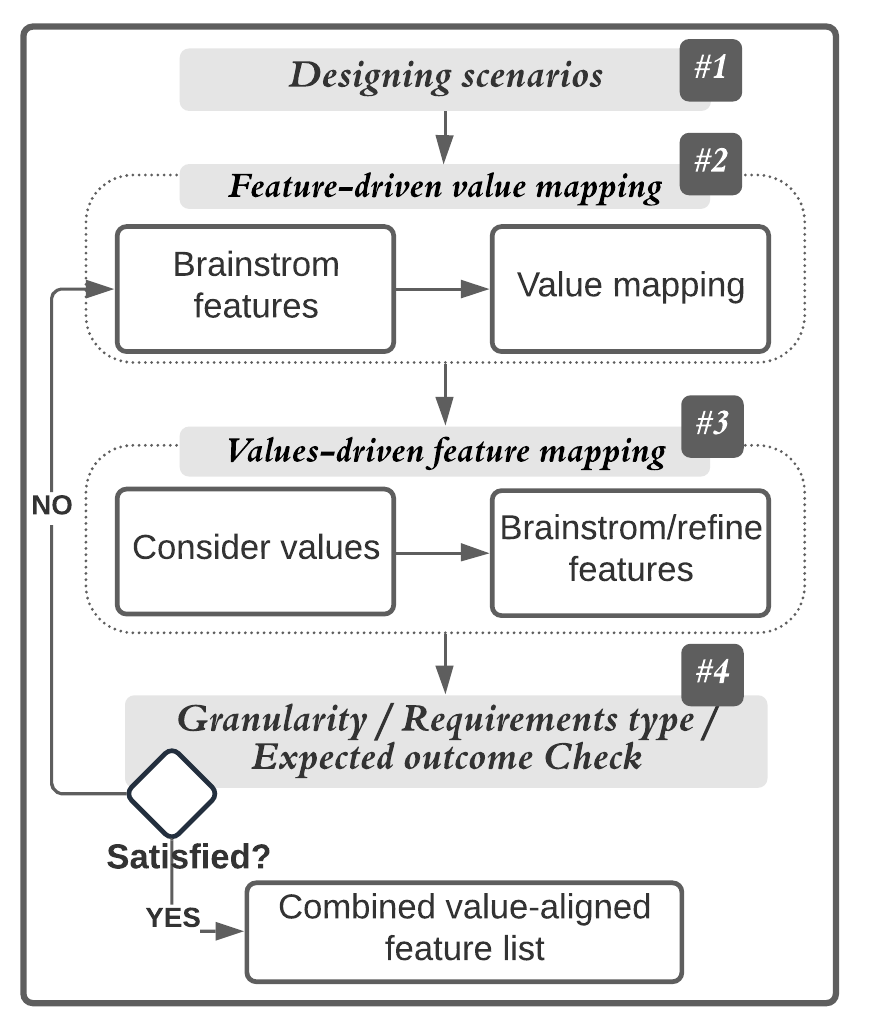}
	\caption{Scenario-based Values Elicitation -- a four-step process (\# denotes step number)}
	\label{fig:fiveStep}
\end{figure}

Based on the results and our experience, we propose a \textit{scenario-based values elicitation} process as a practical implication and takeaway of this study (see Fig.~\ref{fig:fiveStep}).

\textit{Step 1: Designing scenarios.} This step involves some members of the software team coming up with a scenario that captures the \textit{standard requirements} of the software being designed. This is ideally done by the manager, business analyst, or a \textit{values champion}~\cite{hussain2021can}, ideally in consultation with the product owner or customer representative. Our scenario \textit{WeCare} in Fig.~\ref{fig:wecare} can be consulted as a guide for this purpose. The team may also wish to make use of personas to further develop the scenario.\\
     
\textit{Step 2: Feature-driven value mapping.} The step follows the bottom up approach of coming up with features for the scenario, followed by mapping them to the Schwartz' values model. \\
    
\textit{Step 3: Values-driven feature mapping.} Once the identified features are mapped to values, the next step is to follow a top-down approach to ensure a good coverage across the value categories. To do this, the team considers the values in the Schwartz model and come up with more features, or refine the identified features, to align with the value categories. While it may seem intuitive to try and achieve a good coverage across all values, the software being designed may be naturally inclined to mapping with some values over others, so it may not be a good idea to force even coverage.\\
    
\textit{Step 4: Granularity for implementation.} The last step is to check the granularity, feature type and expected outcome of the feature. In terms of granularity, as discussed in Findings~(Section~\ref{sec:Findings}), features may be found in between highly abstract level, i.e., closer to human values and fine-grained level, i.e., closer to implementation. We propose this step as a decision point, where practitioners may go back to brainstorm further to make the suggested feature closer to \textit{task level}, thus closer to operationalization of human values. Similarly, practitioners may fine-tune the suggested features in iterations until satisfied for requirements type (Functional Requirements and NFRs) and expected outcome (in-scope and out-scope features). This step might be helpful to identify the indirect stakeholders (like the public in WeCare app) of the scenario. 

Another approach could be to start \textit{Step 2} with more guidance and structure -- than we did in the survey -- asking the participants to suggest the features with enough details so that implementation details can be drawn from them and associated \textit{tasks} can be written up. The choice of approach (open format vs structured format) depends on who is participating. For example, it may be possible to work on a fine-grained task level if developers were participating. However, managers, customers, and business analysts and other people in non-technical roles are more likely to express features as user stories or themes/general guidelines. Generally, a mix of roles is recommended to achieve healthy discussions of values and optimal mapping at the practical implementation level. 
\section{Limitations and Threats to Validity}
\label{sec:ThreatoValidity}
\textit{Internal Validity:} 
The structure and questions of the scenario-based survey may have introduced some threats to the internal validity. All the authors reviewed the survey questions and agreed upon the questions. Furthermore, we improved the survey questions based on feedback that we received in the pilot phase. More specifically, the comprehensibility of the scenario used in the survey might be a question since it was about a specific social group, i.e., homeless people in Australia. It is unknown to what extent each participant was familiar with such context. We used the pilot phase to understand the level of comprehensibility of the scenario introduction and improved the scenario description to be easier to follow. To further support, a hyperlink was added where necessary to that has access to the scenario (Fig.~\ref{fig:wecare}).

The subjective judgment in the process of mapping the suggested feature descriptions to human values may also become a source of threat to internal validity. We mitigated this potential bias by employing three analysts who worked individually during the coding process as prescribed in \textit{investigator triangulation} \cite{bick2017coordination}. Additionally, these analysts also had a similar understanding of human values in SE/RE.



\textit{Construct Validity:}  Possible threats to construct validity may arise from the participants' and the analysts' understanding of human values. The human values theory we used might have been completely new to the survey participants, which may have led to misinterpretations of values. We identified the lack of familiarity with the provided human values definitions and examples among the participants in the pilot phase. In the main study, our strategy to help participants better understand human values was to add hyperlinks, where necessary, to an external document. The external document contained detailed information about Schwartz's theory, including the values circular model.
As for the analyst, they all had a decent understanding of human values and had research experience on human values in software.

\textit{External Validity:} Given the number of survey participants, we accept that the findings and conclusions may not be applicable to the entire global RE community. Nevertheless, the participants came from 15 different job roles and had a range of years of experience (less than a year to 25 years), which can be a reasonable representation of the RE community.

\section{Conclusion and Future Work}
\label{sec:Conclusion}

The demand for software that reflects human values is increasing, and Requirements Engineering (RE) has a crucial responsibility of designing software features to demonstrate desired human values. This scenario-based survey contributes to identifying the impact of explicit consideration of human values during RE activities. Our survey attracted 56 participants who are mainly involved in RE activities in their day-to-day life.  
The results suggest practitioners may confidently consider human values during RE activities as it allows them to (i) identify values related to a given scenario, (ii) associate purpose with the features they develop, (iii) be creative as well as draw ideas from life experiences, and (iv) build connections between human values and software features.

Further, we find human values alignment with features could be done effectively using either \textit{feature-driven value mapping} (brainstorm features, then map to values ) or \textit{values-driven feature mapping} (triggering with values to brainstorm features).
Finally, scenario-based values elicitation – a four-step takeaway process for practitioners to use scenarios to elicit values and develop a value-aligned feature list for a given scenario(s). 

In future work, we will address several points. First,
as discussed, we will look for potential guidelines, tools and techniques that can handle explicit values consideration in RE and support the scenario-based values elicitation process to be effectively practiced in the industry. 
Second, we plan to continue to research on \#\textit{ValuesTrade-offs} and potentially extend the scenario-based values elicitation process to accommodate and evaluate values trade-offs. Third, we will explore ways to extend and trace the explicit consideration of human values throughout subsequent steps in the requirements engineering tasks. Finally, we will further improve the scenario-based survey technique reflecting on learning from this study to be an effective research tool.
\\

\bibliography{references.bib}
\bibliographystyle{IEEEtran}
\end{document}